\documentclass[prl,twocolumn,showpacs,amsmath]{revtex4}
\usepackage{graphicx}
\usepackage{color}

\newcommand{\bl}{\begin{linenomath*}}
\newcommand{\el}{\end{linenomath*}}
\newcommand{\E}{\mathcal{E}}

\newcommand{\be}{\begin{equation}}
\newcommand{\ee}{\end{equation}}
\newcommand{\bea}{\begin{eqnarray}}
\newcommand{\eea}{\end{eqnarray}}
\renewcommand{\vec}[1]{\mathbf{#1}}

\newcommand{\D}{\hat\Psi}
\newcommand{\B}{\hat\Phi}

\begin{document}

% \setpagewiselinenumbers
%\modulolinenumbers[5]
%\linenumbers

%%%%%%%%%%%%%%%%%%%%%%%%%%%%%%%%%%%%%%%%%%%%%%%%%%
\title{Effective magnetic fields for stationary light}
%%%%%%%%%%%%%%%%%%%%%%%%%%%%%%%%%%%%%%%%%%%%%%%%%%

\author{J. Otterbach$^{1}$, J. Ruseckas$^{2}$, R.G. Unanyan$^{1}$, G. Juzeli\=unas$^{2}$, M. Fleischhauer$^{1}$}
\affiliation{$^{1}$Fachbereich Physik and research center OPTIMAS, Technische Universit\"at Kaiserslautern, 67663 Kaiserslautern, Germany\\
$^{2}$ITPA, Vilnius University, 01108 Vilnius, Lithuania}

%\author{James R. Anglin}
%\affiliation{Fachbereich Physik, Technische Universit\"at Kaiserslautern, 
%D-67663 Kaiserslautern, Germany}

\date{\today}

%%%%%%%%%%%%%%%%%%%%%%%%%%%%%%%%%%%%%%%%%%%%%%%%%%
\begin{abstract}
We describe a method to create effective gauge potentials for stationary-light polaritons. When stationary light is created in the interaction with a rotating ensemble of coherently driven double $\Lambda$ atoms, the equation of motion is that of a massive Schr\"odinger particle in a magnetic field. Since the effective interaction area for the polaritons can be made large, degenerate Landau levels can be created with degeneracy well above 100. This opens the possibility to study the bosonic analogue of the fractional quantum Hall effect for interacting stationary-light polaritons.
\end{abstract}
%%%%%%%%%%%%%%%%%%%%%%%%%%%%%%%%%%%%%%%%%%%%%%%%%%%

\pacs{41.20.Jb, 42.50.Gy, 42.50.Ct}

\maketitle
%%%%%%%%%%%%%%%%%%%%%%%%%%%%%%%%%%%%%%%%%%%%%%%%%%%
%%%%%%%%%%%%%%%%%%%%%%%%%%%%%%%%%%%%%%%%%%%%%%%%%%%

%%%%%%%%%%%%%%%%%%%%%%%%%%%%%%%%%%%%%%%%%%%%%%%%%%%
%%%%%%%%%%%%%%%%%%%%%%%%%%%%%%%%%%%%%%%%%%%%%%%%%%%
% \section{introduction}
%%%%%%%%%%%%%%%%%%%%%%%%%%%%%%%%%%%%%%%%%%%%%%%%%%%
%%%%%%%%%%%%%%%%%%%%%%%%%%%%%%%%%%%%%%%%%%%%%%%%%%%

One of the outstanding and challenging problems of many-body physics is the understanding of  strongly correlated quantum systems. With the technological advances of atomic physics and quantum optics over the last decades a number of new model systems based on cold atoms emerged which allow an experimental study with unprecedented precision and control \cite{Bloch-RMP-2008}. Recently it has been suggested to consider 
quasi-particles arising due to the strong light-matter coupling as an alternative.  It has been predicted in \cite{Chang-Nature-Phys-2008} that slow-light polaritons \cite{Fleischhauer-PRL-2000} in a nonlinear fiber would undergo a crystallisation transition similar to the Tonks-Giradeau transition \cite{Tonks-Girardeau}. The dynamics of strongly interacting polaritons  was considered  in a coupled cavity array \cite{Hartmann-NPhys-2006}, realizing the Jaynes-Cummings-Hubbard model. Furthermore a mechanism to induce Bose-Einstein condensation  \cite{Fleischhauer-PRL-2008} of stationary light polaritons \cite{Zimmer-PRA-2008} was proposed and analyzed.

In this paper we show that it is possible to create effective gauge potentials for stationary-light polaritons along with non-zero effective magnetic fields. This extends previous related proposals for the generation of gauge potentials for atoms \cite{Juzeliunas}. It opens up the possibility to study a variety of single- and many-particle effects in magnetic fields, such as the action of a Lorentz force or, in the presence of interactions, the bosonic analogue of the fractional quantum Hall effect \cite{Joliceour} for stationary light. The achievable strength of the magnetic field is comparable to the case of atoms \cite{Juzeliunas}, but  the polaritons have a number of advantages. Stationary-light polaritons \cite{Zimmer-PRA-2008,Moiseev-PRA-2006} emerge in the interaction of a pair of counter-propagating light fields with a double-$\Lambda$ atomic system driven by a pair of counter-propagating control laser. They behave as Schr\"odinger particles with an effective mass that can be adjusted by the control fields. Spatial confinement to lower dimensions can be achieved by simple wave-guide and resonator techniques and the effective temperature can be controlled.

In the following we show how a non-zero effective magnetic field can be generated for the stationary light using a uniformly rotating medium, similar to cold atoms in rotating traps \cite{Bloch-RMP-2008,Viefers-JOP-2008}. In the case of the electrically neutral cold atoms the artificial magnetic field can also be created using two counterpropagating light beams with shifted spatial profiles \cite{Juzeliunas-PRA-2006,Cheneau EPL-2008} or having a transverse dependence of the atomic energy levels \cite{Spielman-2008-archive}. Yet in the present scheme it is the stationary polaritons rather than the atoms that are affected by the gauge field. The underlying mechanism can be attributed to a rotational frequency shift \cite{Birula-PRL-1997,Ruseckas-PRA-2007}.
There have been proposals of creating gauge fields for slow-light via the spatial dependence of the control beams \cite{Marzlin-PRA-2008} or using moving media to induce an Aharonov-Bohm phase  \cite{Leonhardt-JMO-2001} and light drag \cite{Ohberg-PRA-2002,Juzeliunas-PRA-2003} for slow light. In contrast we here discuss the stationary rather than the slow light setup which allows generation of Landau levels with a high degree of degeneracy.

\begin{figure}[t]
\begin{centering}
  \includegraphics[width=0.45\textwidth]{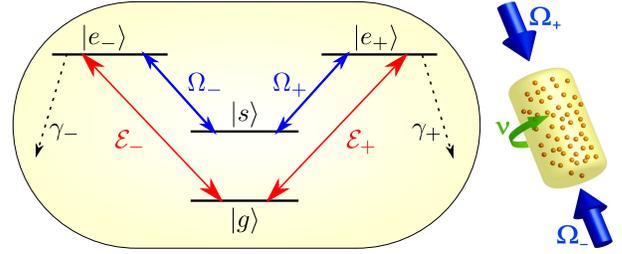}
\end{centering}
\caption{(Color online) The Raman interaction of two counterpropagating control
lasers of Rabi-frequencies $\Omega_{\pm}$ and opposite circular polarization
coupling to the $|s\rangle-|e_{\pm}\rangle$ transitions of a double $\Lambda$
system generates a quasi-stationary pattern of Stokes fields $\mathcal E_{\pm}$, called
stationary light. The stationary dark-state polaritons formed in this
interaction do not decay radiatively and have a quadratic dispersion profile in
all directions.}
\label{fig:double-lambda}

\end{figure}

We consider a four level scheme involving two hyperfine atomic ground states $|g\rangle$ and $|s\rangle$ with magnetic quantum numbers $m=0$, as well as two excited states $|e_{\pm}\rangle$ with $m=\pm1$, as shown in Fig. \ref{fig:double-lambda}. The states are coupled in a closed loop configuration by four light fields with opposite circular polarizations. A pair of counterpropagating control lasers with Rabi-frequencies $\Omega_{\pm}e^{\pm ik_cz}$ drives the transitions $|s\rangle\rightarrow|e_{\pm}\rangle$ to create electromagnetically induced transparency (EIT) for another pair of counterpropagating quantized probe fields $\hat{E}_{\pm}$ coupling the states $|g\rangle$ and $|e_{\pm}\rangle$. This sets up two parallel $\Lambda$ schemes sharing the same ground states. EIT appears if the two-photon resonance is maintained for both $\Lambda$ systems $\omega_{+}^{(p)}-\omega_{+}^{(c)}=\omega_{-}^{(p)}-\omega_{-}^{(c)}=\omega_{sg}$. If the amplitudes of the control fields are equal, $|\Omega_{+}|=|\Omega_{-}|$, a stationary light polariton is formed \cite{Zimmer-PRA-2008,Fleischhauer-PRL-2008}.

Let us introduce field amplitudes $\hat{\mathcal{E}}_{\pm}$ that are normalized
to a number and vary slowly in space and time by $\hat{E}_{\pm}(\mathbf{r},t)=\sqrt{\frac{\hbar\omega}{2\varepsilon_0}}\hat{ \mathcal{E}}_{\pm}(\mathbf{r},t)\exp\{-i(\omega_pt\mp k_pz)\}+\mathrm{h.c.}$ .
Furthermore continuous atomic-flip operators are defined as $\hat{\sigma}_{\mu\nu}(\mathbf{r},t)=\frac{1}{\Delta N}\sum_{j\in\Delta V(\mathbf{r})}\hat{\sigma}_{\mu\nu}^j$, where $\hat{\sigma}_{\mu\nu}^j\equiv|\mu\rangle_{jj}\langle\nu|$ is the flip operator of the $j$-th atom, and the sum is taken over a small volume $\Delta V$ around $\mathbf{r}$ containing $\Delta N$ atoms.

In what follows the probe fields are considered weak. This prevents depletion of the ground state $|g\rangle$. We further assume the medium to rotate uniformly with angular frequency $\nu$ around its symmetry axis, which is the propagating direction of control and probe fields. As a result one arrives at a set of equations for the atomic coherences and the probe beams in the lab frame:
%
%
%%\bl
\begin{align}
i\Bigg(\frac{\partial}{\partial t} & +i\nu\frac{\hat L_z}{\hbar}\Bigg) \hat{\sigma}_{gs} \,=\,
\delta\hat{\sigma}_{gs}-\Omega_{+}\hat{\sigma}_{ge_{+}}-\Omega_{-}\hat{\sigma}_{
ge_{-}}
\label{eq:sigma-gs}
\\ i\Bigg(\frac{\partial}{\partial t} & +i\nu\frac{\hat L_z}{\hbar}\Bigg)\hat{\sigma}_{ge_{\pm}}\, =\,
- i\Gamma_{\pm}\hat{\sigma}_{ge_{\pm}}-\Omega_{\pm}\hat{\sigma}_{gs}\nonumber \\
&\qquad\qquad\qquad\quad-g\sqrt{
n}\,\hat{\mathcal{E}}_{\pm}+i\hat F_\pm,
\label{eq:sigma-ge}
\\ i\frac{\partial}{\partial t} & \hat{\mathcal{E}}_{\pm} \,=\, \left[\mp ic\frac{\partial}{\partial
z}-\frac{c}{2k_p}\nabla^2\right]\hat{\mathcal{E}}_{\pm}-g\sqrt{n}\hat{\sigma}_{
ge_{\pm}}\,\label{eq:probe-field}.
\end{align}
%\el
%
%
Here $\hat L_z$ is the orbital angular momentum of the atoms along the $z$-axis, $n$ is the atom density and $g=\frac{\wp}{\hbar}\sqrt{\frac{\hbar\omega}{2\varepsilon_0}}$ is the common coupling constant of both probe fields with $\wp$ denoting the dipole matrix element. $\Gamma_{\pm}=\gamma_\pm+i\Delta_{\pm}$, where $\gamma_\pm$ is the decay rate of the transitions $|e_{\pm}\rangle-|g\rangle$ and $k_p=\omega_p/c$ is the carrier wavenumber of the probe. The single-photon detunings of the upper states are denoted by $\Delta_{+}$ and $\Delta_{-}$, respectively, and $\delta$ stands for a small two-photon detuning. $\hat F_A$ are Langevin noise operators necessary to preserve the commutation relations. For exponentially decaying variables the noise operators are $\delta$-correlated in time $\langle \hat F_A(t)\hat F_B(t')\rangle\,=\,D_{AB}\delta(t-t')$
and the diffussion coefficients $D_{AB}$ are proportional to the population of the excited states. Since we work in the linear response regime, i.e. the population of the excited states is negligible, we are allowed to disregard the Langevin noise terms.

The adiabatic eigensolution of the coupled Maxwell-Bloch eqs.(\ref{eq:sigma-gs})-(\ref{eq:probe-field}) 
immune to spontaneous decay is the stationary dark-state polariton (DSP) \cite{Zimmer-PRA-2008}
%
%
%\bl
\begin{align}
\D=\cos\theta(\cos\varphi\hat{\mathcal{E}}_{+}+\sin\varphi\hat{\mathcal{
E}}_{-})-\sin\theta\hat{\sigma}_{gs}\,,
\label{eq:DSP-definition}
\end{align}
%\el
%
%
where the mixing angles are defined as $\tan\theta=g\sqrt{n}/\Omega$, with $\Omega^2=\Omega_{-}^2+\Omega_{+}^2$ and $\tan\varphi=\Omega_{-}/\Omega_{+}$.
In the following we assume equally strong and real control fields, i.e. $\Omega_+=\Omega_-$, corresponding to the creation of stationary light polaritons \cite{Fleischhauer-PRL-2008, Zimmer-OC-2006}. 
%%%%%%%%%%%%%%%%%%%%%%%%%%%%
%  Furthermore having in mind applications to quantum Hall physics we assume a bimodal cavity to freeze out the 
% longitudinal motion \cite{Messina-JMO-2003, Rauschenbeutel-PRA-2001}. Thus after dropping all derivatives along the 
% $z$-axis, 
%%%%%%%%%%%%%%%%%%%%%%%%%%%%%
Thus the equation of motion for the DSP is given by
%  [\textbf{MF: we argue at the beginning that we consider
% stationary light, if we use a cavity to be 2D there is no need for that however. I suggest to stick to the
% full 3D description. We are not considering quantum Hall physics in this paper. Johannes can you
% make the corresponding changes, i.e. put in the z derivatives and the longitudinal mass again ??}
%\sin 2\varphi     -c\sin\theta\cos\theta\cos 2\varphi\frac{\partial}{\partial z}
%+c\cos^2\theta\cos 2\varphi\frac{\partial}{\partial z}
%\bl
\begin{align}
&\Big[\frac{\partial}{\partial t} +i\nu\sin^2\theta\frac{\hat L_z}{\hbar}-i\frac{v_g}{2k_p}\nabla_\perp^2\Big]\D -c\cos\theta \frac{\partial}{\partial z}\B_1\, \nonumber \\
-& \Big[i\nu\sin\theta\cos\theta\frac{\hat L_z}{\hbar} + i\frac{c}{2k_p}\sin\theta\cos\theta\nabla_\perp^2\Big]\hat\Phi_2 \nonumber \\ 
=&\, i\delta\left(\sin^2\theta\D-\sin\theta\cos\theta\B_2\right),
	\label{eq:EoM-DSP}
\end{align}
%\el
%
%
where $v_g=c\cos^2\theta$ is the EIT group velocity and $\nabla_\perp=\left(\partial_x,\partial_y\right)^T$. Here $\B_1=-\sin\varphi\E_++\cos\varphi\E_+$ and $\B_2=\sin\theta(\cos\varphi\hat{\mathcal{E}}_{+}+\sin\varphi\hat{\mathcal{
E}}_{-})+\cos\theta\hat\sigma_{gs}$ are the other eigensolution of eqs. (\ref{eq:sigma-gs})-(\ref{eq:probe-field}) whose equations of motion after elimination of the excited states read
%
%-c\cos 2\varphi \frac{\partial}{\partial z}
%\bl
\begin{align}
 &\Big[\frac{\partial}{\partial t}-i\frac{c}{2k_p}\nabla_\perp^2\Big]\B_1
-c\cos\theta\frac{\partial}{\partial z}\D-c\sin\theta\frac{\partial}{\partial z}\B_2\nonumber\\
=&-\frac{g^2n}{\Gamma}\B_1,
\end{align}
and
\begin{align}
 &\Big[\frac{\partial}{\partial t}+i\nu\cos^2\theta\frac{\hat L_z}{\hbar}-i\frac{c}{2k_p}\sin^2\theta\nabla_\perp^2\Big]\B_2\, \nonumber \\
-&\Big[+i\nu\sin\theta\cos\theta\frac{\hat L_z}{\hbar} +i\frac{c}{2k_p}\sin\theta\cos\theta\nabla_\perp^2 \Big]\D \nonumber \\
-&c\sin\theta\frac{\partial}{\partial z}\B_1 \nonumber \\
=&\,-\frac{g^2n+\Omega^2}{\Gamma}\B_2\approx-\frac{g^2n}{\Gamma}\B_2.
\label{eq:EoM-BSP}
\end{align}
%\el
%
%
Here we put $\Gamma_+=\Gamma_-=\Gamma$ and neglected terms containing $\delta$ because of the EIT condition $\Gamma\delta\ll\Omega^2$. The last line holds as long as $g^2n\gg\Omega^2$, which is easily fulfilled for usual slow- and stationary-light experiments \cite{Hau-Nature-1999, Bajcsy-nature-2003}.
Adiabatic elimination of $\B_2$ in eq. (\ref{eq:EoM-BSP}) and subsequent subsitution into eq. (\ref{eq:EoM-DSP}) results 
in the following equation for the DSP which is correct up to second order in non-adiabatic corrections:
%
%
%\bl
\begin{align}
i\hbar\frac{\partial}{\partial t}\D\,=&\,\left[\frac{1}{2m_\parallel}\hat{p}_z^2+\frac{1}{2m_\perp}\left(\hat{\vec p}_\perp +\vec{A}\right)^2+U\right]\D \nonumber\\
&-i\frac{\Gamma^\perp_\text{rot}}{\hbar}\hat L_z^2\D-i\hbar D^\parallel_\text{diff}\frac{\partial^2}{\partial z^2}\D.
	\label{eq:EoM-DSP-minimal-coupling-form}
\end{align} 
%\el 
%
%
Eq. (\ref{eq:EoM-DSP-minimal-coupling-form}) represents a Schr\"odinger equation for a particle with an effective tensorial mass with minimal-coupling to an artificial gauge field. The masses are given by $m_\parallel= \hbar\gamma/2 v_g L_\text{abs}\Delta $ and $m_\perp=\hbar k_p/v_g$, respectively. $L_\text{abs}=c\gamma/g^2n$ defines the resonant absorption length in absence of EIT. The gauge field can be expressed as
%
%
%\bl
\begin{align}
\vec{A}\,&=\,m_\perp (\nu\vec{e}_z\times\vec r_\perp)\sin^2\theta.
%A_\varphi\vec{e}_\varphi,\quad A_\varphi\,=\,m_\perp\nu\rho\sin^2\theta	
	\label{eq:VectorPotential}
\end{align}
%\el
%
%
$\vec r_\perp=(x,y)^T$ is the radius vector from the axis of rotation and the corresponding momentum operator is given by $\hat{\vec p}_\perp=-i\hbar\nabla_\perp$. The scalar potential reads
%
%
%\bl
\begin{align}
	U\,=\,-\frac{1}{2}m_\perp\nu^2\rho^2\sin^4\theta+\hbar\delta\sin^2\theta
	\label{eq:ScalarPotential}
\end{align}
%\el 
%
%
where $\rho=\|\vec r_\perp\|$. Choosing a proper two-photon detuning, by e.g. spatially varying Zeeman or Stark shifts, the antibinding part of the scalar potential can be compensated. Finally the rotation induced and longitudinal diffusion rates $\Gamma^\perp_\text{rot}$ and $\Gamma^\parallel_\text{diff}$ appearing in eq.(\ref{eq:EoM-DSP-minimal-coupling-form}) are
%
%
%\bl
\begin{align}
 \Gamma^\perp_\text{rot}\,&=\,\frac{L_\text{abs}}{v_g}\nu^2\sin^2\theta\cos^4\theta\left(1+i\frac{\Delta}{\gamma}\right),
 \label{eq:angular-diffusive-lossrate} \\
D^\parallel_\text{diff}\,&=\,v_g L_\text{abs}. 
 \label{eq:diffusion-rate}
\end{align}
%\el 
%
%
The real part of $\Gamma^\perp_\text{rot}$ describes azimutal diffusion associtaed with loss and the imaginary part a corresponding correction to the mass. $D^\parallel_\text{diff}$ is responsible for a diffusive behavior along the original propagation axis of the light.

From eq. (\ref{eq:VectorPotential}) we compute the magnetic field to be
%
%
%\bl
\begin{align}
 \vec B=\nabla \times \vec A=2m_\perp\nu\sin^2\theta\; \vec e_z.
 \label{eq:magneticField}
\end{align}
%\el 
%
%
This expression is identical to the effective magnetic field $B_\text{eff}=2 m_\text{at} \nu$ created by rotating cold gases \cite{Viefers-JOP-2008} except for the factor $\sin^2\theta$ and the substitution of atomic mass with the mass of the quasiparticles. This can be understood as follows: The particles feeling the magnetic field are the polaritons rather than the atoms. According to eq. (\ref{eq:DSP-definition}) these particles are a superposition of photonic and matter excitation and only the matter component, proportional to $\sin^2\theta$ and with effective
mass $m_\perp$, is subject to rotations. For $\sin^2\theta=0$ there is no coupling between light and medium and thus no gauge potential emerges for the polariton which in this case is just the electromagnetic field.

From eq.(\ref{eq:magneticField}) we obtain magnetic length and filling
%
%
%\bl
\begin{align}
 L_\text{mag}^2\,&=\,\frac{\hbar}{B}\,=\,\frac{1}{4\pi}\lambda R\frac{v_g}{v_\text{rot}} \label{eq:MagneticLength}\\
 \nu_\text{filling}\,&=2\pi n_\Psi L_\text{mag}^2\,=\,\frac{1}{2}N_\Psi\frac{\lambda}{R}\frac{v_g}{v_\text{rot}}. 
\label{eq:FillingFactor}
\end{align}
%\el
%
%
Here we introduced the rotation velocity $v_\text{rot}=\nu R$ of the medium at its circumference $\rho_\text{max}=R$, $N_\Psi$ is the number of DSPs  and $\lambda=2\pi/k_p$ is the wavelength of the probe field.

The adiabaticity condition imposed by rate (\ref{eq:angular-diffusive-lossrate}) reads ${\rm Re}\left[\Gamma^\perp_\text{rot}\right]\stackrel{!}{\ll}\omega_c$,
where the cyclotron frequency $\omega_c\equiv B/m_\perp=2\nu\sin^2\theta$ sets the time scale of the relevant physics. Evaluating this we find the following condition
%\bl
\begin{align}
\frac{v_\text{rot}}{v_g}\ll\frac{R}{L_\text{abs}}\frac{1}{\cos^4\theta}\; \Leftrightarrow\;\nu\,\ll\frac{v_g}{L_\text{abs}}\frac{1}{\cos^4\theta}
 \label{eq:adiabaticityCondition2}.
\end{align}
%\el
This is easily fulfilled, since for typical group velocity $v_g\sim 10^3$ m/s and absorption length $L_\text{abs}\sim 1$cm the resulting mixing angle is $\cos^2\theta=v_g/c\approx10^{-5}$ and thus either right hand side of eq.(\ref{eq:adiabaticityCondition2}) is large.

Analogously we obtain a lower boundary for the rotation frequency by the adabaticity condition from rate (\ref{eq:diffusion-rate}) leading to $\omega_c \stackrel{!}{\gg}\frac{D^\parallel_\text{diff}}{L_p^2}$ which yields
\begin{align}
 \nu \gg\frac{1}{2}\frac{v_g}{L_\text{abs}}\left(\frac{L_\text{abs}}{L_p}\right)^2.
 \label{eq:adiabaticityCondition3}
\end{align}
Here $L_p$ stands for the characteristic length scale of the stationary DSP along the $z$-axis. The ratio $v_g/L_\text{abs}$ is the inverse time scale it takes a photon to travel one absorption length. Condition (\ref{eq:adiabaticityCondition3}) demands that the rotation frequency is larger than this inverse decay time in order to see interesting physics.

There are numerous experimental systems which seem suitable for the implementation of the above suggested scheme. The choice of the systems is guided by the possibility to create strong and controllable  interactions between individual polaritons to eventually explore effects such as the bosonic fractional quantum Hall effect. In addition to contact interactions of cold atoms in a trap it is possible to create long-range interactions by exploiting the interaction properties of the matter component of the polaritons.

\textit{Bulk materials. -- } A straightforward realization is to use rotating bulk media. Rare-earth-ion doped glasses exhibit long coherence times up to $T_2=82$ms \cite{Fraval-PRL-2004} and are proposed to be used for quantum computation \cite{Ohlsson-OC-2002}. Interactions can be created via dipole-dipole interactions between hyperfine states of different principal electronic states or via photonic nonlinearities. A table listing several experimental data of rare-earth-ion dopants can be found in \cite{McAuslan-arXiv-2009}.
Also n-doped semiconductors as GaAs \cite{Wang-OC-2007} can create strong interactions by exploiting the Coulomb interaction between excitons, which constitute the matter component of the polaritons. However coherence times are only of the order of several ns and thus too short. It is believed that these short lifetimes are caused by the coupling of the electron spin to the nuclear spins. A possible way to overcome this obstacle would be to use Si-based structures as a host, which does not posses a nuclear spin, and show coherence times up to 60 ms \cite{Tyryshkin-PRB-2003}.

\textit{Rotating optical lattices. -- } In \cite{Williams-OE-2008} the creation of a rotating optical lattice with rotation frequencies up to several kHz is reported. This could be exploited to uniformly rotate cold atoms or polar molecules in the gas phase as a bulk medium and at the same time taking advantage of the high-precision techniques developed in this field. To create long-range interactions one may think of using Rydberg atoms or polar molecules. Recent works on Rydberg atoms report of the successful creation of Rydberg excitations in a BEC \cite{Heidemann-PRL-2008} with blockade radii of $r_b=5.4\mu$m. The lifetime of these systems is of the order of $\sim 100\mu$s \cite{Raitzsch-NJP-2009}. Alternatively one could think of loading the optical lattice with polar molecules. There have been suggestion to create single-photon nonlinearities with these molecules \cite{Yelin-PRA-2006} and investigations about the experimental feasibility \cite{Kuznetsova-PRA-2008} stating that lifetimes of about $\sim 1$s are achievable.

It should be noted that cold atoms and polar molecules can also be embedded in solid-state matrices \cite{Andrews-1989}, which can then be physically rotated.

To observe an artificial magnetic field for the DSP we suggest two possibilities.
A first indication would be the observation of a Lorentz force acting on a slow light polariton $ \vec F\,=\,\frac{1}{m_\perp}\langle\hat{\vec p}\rangle\times\vec B$. Shining in a probe beam with a small transverse extent along a propagation axis shifted from the axis of rotation will result in a small deflection of the incoming light pulse from its initial direction \cite{Padgett-OL-2006}. The deflection angle is given by $\Delta\alpha\,=\,\omega_c \rho \frac{L}{v_g}$, where $\rho$ is the distance of the initial beam axis from the rotation axis.
Using a stationary light setup the outcome strongly depends on the initial mode profile. Creating the stationary DSPs using modes of the probe beams that are not eigenmodes of the angular momentum operator, e.g. higher Hermite-Gaussian modes, and releasing after a time will result in an image rotation. The angle of rotation is directly proportional to the storage time of the probe light inside the medium. If the initial state is a superfluid of the DSPs \cite{Fleischhauer-PRL-2008} the artificial magnetic field leads to the formation of a vortex lattice. The structure of the lattice will be well visible upon releasing the stationary polaritons
providing a convenient means to observe the lattice.

In summary we presented a possible scheme to create artificial gauge fields for photonic quasi-particels, the so-called dark-state polaritons. The size of the resulting effective magnetic field is large enough to create highly degenerate Landau levels. Observation of the artificial fields is possible by turning the stationary light polaritons into slow light polaritons \cite{Zimmer-OC-2006} and detecting the transverse emission profile. We suggested several physical systems which, to our knowledge, seem suitable for the implementation of the above ideas. In the future we will study interactions between polaritons in order to investigate the bosonic fractional quantum Hall effect.

%%%%%%%%%%%%%%%%%%%%%%%%%%%%%%%%%%%%%%%%%%%%%%%%%%%%%

\textit{Acknowledgments. --} This work was supported by the DFG through the GRK 792 and by the Alexander von Humboldt Foundation.

%%%%%%%%%%%%%%%%%%%%%%%%%%%%%%%%%%%%%%%%%%%%%%%%%%%%%

%%%%%%%%%%%%%%%%%%%%%%%%%%%%%%%%%%%%%%%%%%%%%%%%%%%%%%

\end{document}